\documentclass{elsart}

\begin{document}
\begin{frontmatter}

\title{Thermal measurements of stationary nonequilibrium systems:
A test for generalized thermostatistics}

\author{Dami\'an H. Zanette}

\address{Consejo Nacional de
Investigaciones Cient\'{\i}ficas  y T\'ecnicas, Centro At\'omico
Bariloche and Instituto Balseiro, 8400 San Carlos de Bariloche,
R\'{\i}o Negro, Argentina}

\ead{zanette@cab.cnea.gov.ar}

\author{Marcelo A. Montemurro}

\address{Abdus Salam
International Centre for Theoretical Physics, Strada Costiera 11,
34100 Trieste, Italy}

\ead{mmontemu@ictp.trieste.it}

\begin{abstract}
We show that a gas thermometer in contact with a stationary
classical system out of thermal (Boltzmann) equilibrium evolves,
under very general conditions, towards a state characterized by a
L\'{e}vy velocity distribution. Our approach is based on a
kinetic-like equation that applies to a wide class of models for
the system-thermometer interaction. The results clarify the role
of non-exponential energy distributions as possible
generalizations of the Boltzmann distribution for systems where
the usual formulation of thermostatistics may not apply. In
particular, they show that the power-law distributions derived
from Tsallis's nonextensive formalism are irrelevant to the
stationary state of the thermometer, thus failing to give a
consistent description of the system-thermometer equilibrium. We
point out the need of a generalized thermostatistical formulation
able to give a unified frame to L\'evy and Maxwell distributions.
\end{abstract}

\begin{keyword}

Non-equilibrium systems \sep Kinetic theory \sep L\'evy distributions
\sep Generalized thermostatistics
 
\PACS 05.20.-y \sep  05.70.-a \sep 05.20.Dd
\end{keyword}
\end{frontmatter}

L\'evy distributions \cite{MW} have been extensively applied to the 
description and modeling of a wide class of physical processes, 
ranging from anomalous transport in disordered media \cite{Bouchaud} 
and turbulent flows \cite{turbul0}, to phase-space diffusion in 
dynamical systems \cite{phase} and polymer dynamics \cite{polymer}. 
They have also found application in other branches of 
science, such as in biology \cite{b1,b2}. The ubiquity of L\'evy 
distributions in many natural phenomena is a straightforward
consequence of their stable character under summation of random 
variables whose distributions have diverging moments, as shown 
by Paul L\'evy in his generalization of the central limit 
theorem \cite{L}. This essential result gives L\'evy distributions 
the  same status as the Gaussian distribution in the statistical 
description of stochastic processes.   

While most applications of L\'evy distributions deal with 
dynamical and transport processes, we show in this Letter 
that they replace the {\it equilibrium} Maxwell distribution---a 
Gaussian in the velocity variable---in an extended version 
of the scenario of a thermal measurement. We analyze the asymptotic
energy distribution of a thermometer in thermal contact with a
classical system in a nonequilibrium stationary state. While the
nature of this nonequilibrium state is not explicitly specified, we
typically refer to a situation where the system is not isolated and
suffers the effect of external stationary forces that maintain its
energy distribution apart from the prediction of Boltzmann-Gibbs
statistics. It is an essential fact of thermostatistics \cite{callen}
that, if the system is in thermodynamical equilibrium,  its Boltzmann
energy distribution is ``copied'' by the thermometer in such a way
that its average energy per degree of freedom, i.e. its temperature,
becomes equal to that of the system. Now, what properties of the
state of the system are detected by the thermometer in the case that
the system is maintained out of equilibrium?

The motivation of this question is two-fold. First, since all physical
systems are to some extent thermodynamically open and subject to
external influence, the possible effect of these factors in a thermal
measurement constitutes a problem of empirical relevance. In general,
this problem transcends the limits of equilibrium thermostatistics
and calls for a nonequilibrium, dynamical formulation. We choose a
kinetic-like approach which describes a wide class of interaction
models for the thermal contact between the system and a gas
thermometer. Our results show that the thermometer attains a velocity
distribution which is fully determined by the high-energy
distribution of the system, and results to be insensitive to other
details in the system state. In the case where the energy
distribution of the system decays as a power law, as advanced above,
the thermometer velocities approach a L\'{e}vy distribution. This
distribution ``copies'' the singularities associated with divergent
energy moments in the system, and reduces to a Maxwellian when the
average energy is finite.

The second motivation for our question has to do with the possibility
of generalizing thermostatistics to the description of physical
systems where the usual Boltzmann-Gibbs formulation may not apply
\cite{ts2}. Tsallis's nonextensive thermostatistics, for instance, is
claimed to replace the usual formulation for systems where long-range
interactions lead extensivity assumptions to break down, while
containing Boltzmann-Gibbs statistics as a limiting case \cite{ts1}.
In Tsallis's formalism, the Boltzmann exponential distribution is
replaced by a class of power-law functions derived from a variational
principle for a generalized entropy. Whether this formalism is
compatible with the basic facts of equilibrium between systems in
thermal contact can be decided, precisely, by studying the
interaction of a thermometer and a system with such power-law energy
distribution. Our results show that, since L\'{e}vy distributions are
not regarded in this formalism as possible equilibrium states,
Tsallis's thermostatistics fails to give a consistent picture of
thermodynamical equilibrium.

Consider a system in a stationary state which does not necessarily
coincide with Boltzmann equilibrium. We assume that the state of the
system is macroscopically characterized by an energy distribution
$F_0(\epsilon)$. The system interacts with a thermometer consisting
of an ensemble of independent particles of mass $m$ moving in a
$d$-dimensional domain. Within this domain, the distribution of
particles is supposed to be spatially homogeneous and isotropic in
velocities, such that the associated distribution function
$F(\epsilon,t)$ depends on the energy $\epsilon$ and the time $t$
only. The interaction between system and thermometer is assumed to
fulfill the following conditions. (i) Interaction events are time
localized and infrequent, such that the typical time between
interactions is large as compared with the relaxation time of the
energy distribution of the system toward $F_0(\epsilon)$. This
insures that (a) at each event, the system is found at its stationary
state and (b) any correlation between the states of system and
thermometer, created by the interaction, dies out before the next
event takes place. (ii) The amount of energy interchanged at each
event is small as compared with the total energy of both the system
and the thermometer.

Within these conditions, it is possible to describe the evolution
of $F(\epsilon,t)$ by means of a kinetic-like equation. On the other
hand, as a consequence of condition (i), no evolution equation is
required for the system distribution. To represent the interaction
between system and thermometer we choose a very generic class of
models---closely related to stochastic Maxwell models
\cite{Ernst}---which considerably simplifies the mathematical
treatment of the kinetic problem and, at the same time, does not
imply severe limitations to the physics of the interaction. In fact,
these models include interactions where momentum conservation can
either hold or be violated, and admit arbitrary angular dependence in
the associated cross sections. They assume that the interaction rate
is independent of the energy and that energy itself is conserved. The
corresponding evolution equation for $F(\epsilon,t)$ stands for the
balance between events where, due to interaction with the system, the
thermometer reaches or abandons its states of energy $\epsilon$. Such
events yield, respectively, positive and negative contributions to
$\partial_t F$. The equation is more conveniently written in the
Fourier representation, for the Fourier-transformed distribution
\begin{equation}
\Phi(u,t)\equiv \phi (k,t)= \int  \exp(-i {\bf k}\cdot {\bf v})
f(v,t) d v^d , \label{Fourier}
\end{equation}
with $u=k^2/2m$. Here,
\begin{equation}
f(v,t)= \frac{\Gamma(d/2) }{2 \pi^{d/2}} mv^{2-d} F(mv^2/2,t)
\end{equation}
is the (isotropic) velocity distribution associated with
$F(\epsilon,t)$ \cite{Ernst}. Due to the normalization of
$F(\epsilon)$, we have $\Phi(0,t)=1$ for all
$t$. With an analogous definition for $\Phi_0(u)$ in terms of
$F_0(\epsilon)$, the Fourier-transformed evolution equation reads
\cite{Ernst}:
\begin{equation} \label{kin}
\partial_t   \Phi(u,t) = \int_0^1 w(s) \Phi_0(s u)
\Phi[(1-s)u,t] ds-\Phi(u,t).
\end{equation}
The kernel $w(s)$ characterizes the interaction, and satisfies the
normalization $\int_0^1 w(s) ds = 1$. When necessary, we assume
that the moments
\begin{equation}
w_\mu = \int_0^1 s^\mu w(s) ds
\end{equation}
are well defined.

The  above  condition  (ii)  imposes  that  $w(s)$  is appreciably
different from zero only  for $s\approx 0$. Taking  this condition
into account, we write $\Phi[(1-s)u,t] \approx \Phi(u,t) -su
\partial_u \Phi(u,t)$, to obtain an approximate form of
Eq.~(\ref{kin}). Its solution  can  be  shown  to asymptotically
approach the stationary distribution
\begin{equation}
\Phi(u) = \exp \left[ \frac{1}{w_1}\int_0^u \frac{du'}{u'}
\int_0^1 w(s) \ln \Phi_0(su') ds  \right].
\end{equation}
Since $w(s)\approx 0 $ except for $s \approx 0$ the asymptotic
solution $\Phi(u)$ is fully determined by the behavior of
$\Phi_0(u)$ close to $u=0$. We pay particular attention to the
special case where, for $u\approx 0$,
\begin{equation} \label{beta}
\Phi_0 (u)\approx 1 - \alpha_0 u^\nu,
\end{equation}
with $\alpha_0>0$ and $0<\nu\le 1$. For $\nu \ne 1$, this form
corresponds to an energy distribution with a power-law tail for
high energies, $F_0(\epsilon) \sim \epsilon^{-\nu-1}$. While $F_0$
is still normalized, its first moment---namely, the average
energy---diverges. In the limit $\nu=1$ a finite average energy is
recovered and the long high-energy tail is lost. For $\Phi_0(u)$
as in Eq.~(\ref{beta}), the stationary Fourier distribution for
the thermometer is
\begin{equation}
\Phi (u) = \exp( -\alpha u^\nu ),
\end{equation}
with $\alpha=\alpha_0 w_\nu/\nu w_1$. Note that, for $0<\nu<1$,
$\Phi(u)$ reproduces the singularity of $\Phi_0 (u)$ at $u=0$.
The corresponding Fourier-transformed velocity distribution, $\phi(k)=
\Phi(k^2/2m)$ [see Eq.~(\ref{Fourier})], reads
\begin{equation} \label{Levyd}
\phi (k)= \exp (-b k^\gamma),
\end{equation}
with $b=\alpha/(2m)^\nu$ and $\gamma=2\nu$ ($0<\gamma\le 2$). Note
that, in the velocity representation, this corresponds to a
L\'{e}vy distribution for each velocity component. In fact, the
Fourier transform of, say, $f(v_x)=\int dv_ydv_z\dots f(v)$ is
$\phi(k_x,0,0,\dots)=\exp (-b |k_x|^\gamma)$. Though, in general,
it is not possible to write explicitly the corresponding form of
$f(v_x)$, it is known that for large $|v_x|$ and $\gamma \ne 2$ it
behaves as $f(v_x) \sim  |v_x|^{-\gamma -1}$ \cite{MW}. For 
$\gamma=2$, $f(v_x)$ is a Gaussian distribution.

We stress that the result (\ref{Levyd}) for the Fourier-transformed
velocity distribution of the thermometer is independent of the
detailed form of the distribution $F_0(\epsilon)$ of the system at
low energies. As long as condition (ii) is satisfied, $\phi(k)$ is
completely determined by the power-law high-energy tail of
$F_0(\epsilon)$.

Let us analyze these results in the few cases where
the velocity distribution can be explicitly obtained. We first
consider the situation where the average energy of the system
under study is finite, $E=\int \epsilon F_0 (\epsilon) d\epsilon 
 < \infty$.
Under this condition we have, in Eq.~(\ref{beta}), $\alpha_0=2E/d$
and  $\nu=1$.  This  gives,  for  the Fourier-transformed velocity
distribution of the thermometer,  the Gaussian $\phi(k) =  \exp(-E
k^2/md)$. In the velocity representation we get the Maxwellian
\begin{equation} \label{Max}
f(v) = \left(\frac{md}{4\pi E}\right)^{d/2} \exp\left(-
\frac{md}{4E} v^2\right).
\end{equation}
The average value of the kinetic energy over this distribution is,
as may be expected, $\langle mv^2/2 \rangle =E$. In other words,
the stationary velocity distribution of the thermometer is a
Maxwellian whose average energy per particle coincides with that
of the system. This result includes of course the case where the
system is itself in thermodynamical (Boltzmann) equilibrium at
temperature $T$, for which $E \propto T$.

For $\nu=1/2$, the energy distribution of the system decays as
$F_0(\epsilon) \sim \epsilon^{-3/2}$ and, consequently, the average
energy diverges. The Fourier-transformed velocity distribution of the
thermometer results to be $\phi(k)= \exp(-bk)$, with $b=\alpha_0
w_{1/2} \sqrt{2/m w_1^2}$. Taking into account its units, we write
$b=p_0/m$, where $p_0$ is a characteristic linear momentum given by
the energy distribution $F_0(\epsilon)$ and the specific form of the
interaction kernel $w(s)$. In velocity space, each component $v_i$
is distributed according to a Cauchy distribution,
\begin{equation}
f(v_i)=\frac{mp_0}{\pi}\frac{1}{p_0^2+m^2v_i^2} .
\end{equation}
Note that, here, the relevant dynamical variables in the
thermometer distribution are the momentum components $mv_i$ and,
accordingly, the relevant parameter is the momentum $p_0$.
Compare with the result for $\nu=1$, Eq.~(\ref{Max}), where the
distribution is a function of the energy and the relevant
parameter is $E$.

The present results shed light on the role of non-Boltzmannian
energy distributions in defining the stationary states of
out-of-equilibrium interacting systems. In particular, they show
that, under the above conditions (i) and (ii), the Gaussian
stationary velocity distribution of an ensemble of independent
particles interacting with a non-equilibrium stationary system
results to be a special instance of the more general situation
where the velocity components have L\'{e}vy distributions. We
stress that, despite the fact this conclusion is limited by the
validity of the assumptions (i) and (ii), these conditions are
fully compatible with the paradigm of thermal measurement
\cite{callen} and therefore refer to a realistic, experimentally
accessible situation.  

As advanced above, our conclusions are particularly
significant in the evaluation of generalized formulations of
thermostatistics as plausible descriptions of physical systems to
which the usual Boltzmann-Gibbs theory may not apply. We focus
our attention on Tsallis's formulation, which has motivated large
amounts of work over the last two decades \cite{ts2}. Within
Tsallis's statistics, canonical probability distributions are
obtained from maximization of a generalized entropy \cite{ts1}
\begin{equation}
S_q=-\frac{1-\sum_i p_i^q}{1-q},
\end{equation}
where $p_i$ is the probability of state $i$, and the sum runs over
all the states  of the system.   The parameter $q$  quantifies the
nonadditivity  of   $S_q$,  and   is  therefore   a  measure    of
nonextensivity   in   the   system   under   consideration.    The
maximization process is subject to the generalized constraint
\cite{renorm}
\begin{equation} \label{Eq}
E_q=\frac{\sum_i \epsilon_ip_i^q}{\sum_ip_i^q},
\end{equation}
with $\epsilon_i$ the energy of state $i$, and leads to the
probability distribution
\begin{equation} \label{qexp}
p_i \equiv p (\epsilon_i) = Z_q^{-1}[1-(1-q)\beta_q
(\epsilon_i-E_q)]^{1/(1-q)},
\end{equation}
where $\beta_q$ is a (``renormalized'' \cite{renorm}) Lagrange
multiplier, analogous to the inverse temperature, and $Z_q$ is  a
normalization constant, analogous  to the partition function.
Boltzmann-Gibbs formulation is  recovered for $q\to 1$. Note
that, for systems whose density of states behave as $\rho
(\epsilon) \propto \epsilon^\sigma$---a class which includes most
classical systems---the probability $p(\epsilon)$
corresponds to a Fourier-transformed energy distribution of the
form of Eq.~(\ref{beta}), with $\nu=-\sigma+(q-2)/(1-q)$.

An attractive feature of Tsallis's generalized thermostatistics
is that it preserves most of the mathematical structure of the
usual theory, including the definition of thermodynamic functions,
Legendre transformations, and even linear nonequilibrium
properties \cite{ts2}. In fact, a large part of the literature on
this topic is devoted to the formal extension of thermodynamical
relations to the generalized formulation. Thus, putting aside its
phenomenological applications to nonequilibrium processes such as
anomalous diffusion \cite{ano} and turbulence \cite{turbul}, the
real {\it tour de force} of Tsallis's formalism lies in the
description of thermal equilibrium for systems where
Boltzmann-Gibbs theory is supposed to fail. This is in fact the
original and most frequently invoked motivation of the
formulation and, consequently, constitutes its genuine source of
validation. Our results clarify whether Tsallis's statistics is
relevant to an aspect that---in spite of its essential role in
thermodynamics---has been scarcely treated in the profuse
literature on Tsallis's theory \cite{abe}, namely, thermal
equilibrium between interacting systems.

Assume to have a system which, due to its nonextensive nature,
exhibits an energy distribution of the form of Eq.~(\ref{qexp}),
as predicted by Tsallis's thermostatistics. Allow furthermore the
system to interact with a thermometer, as  specified above.
Consider first that the nonextensivity  index $q$ and the density
of states of the system are such that $E=\sum_i \epsilon_i p_i
<\infty$. According  to our results, the stationary velocity
distribution of the thermometer is a Maxwellian, corresponding to
a Boltzmann exponential probability for the energy. As expected
for an ensemble of independent particles, the thermometer behaves
as an extensive system ($q=1$).  The value of the parameters in
the thermometer distribution predicted by Tsallis's formalism,
however, results to be wrong. Abe and Rajagopal \cite{abe} have
shown that a formal extension  of equilibrium conditions for two
systems in thermal contact yields, in Tsallis's theory,
$ \beta_q=\beta_{q'}$,  where $q$ and $q'$ are the nonextensivity 
indices corresponding to the two systems \cite{not}. Our results 
show that this relation is generally
not satisfied. In fact, taking $q'=1$ as the nonextensivity index
of the thermometer, Eq.~(\ref{Max}) implies  $\beta_{q'} \propto 
E^{-1}$. On the other hand, $\beta_q$ shows in general a
more complicated dependence on $E$. For instance, it can be
readily shown that, for a system with a density of states $\rho
(\epsilon) \propto \epsilon^\sigma$, one has $\beta_q \propto
E^{(q-1)(\sigma+1)-1}$. Let us also point out that the
generalized average $E_q$ of Eq.~(\ref{Eq}) plays no role in the
parameter that defines the equilibrium distribution of the
thermometer---namely, the inverse temperature $\beta_{q'}$---in
spite of the fact that this generalized average replaces $E$ in
the formal extension of thermodynamics to Tsallis's formalism.

In the case that the average $\sum_i \epsilon_i p_i$ diverges, our
results imply that the thermometer approaches an energy distribution
not belonging to the class of the system distribution,
Eq.~(\ref{qexp}). Thus, no well-defined index $q$ can be assigned to
the thermometer. In this situation, the assumption that Tsallis's
thermostatistics would describe the system-thermometer equilibrium
fails drastically. The only trace of the energy distribution of the
system in the thermometer distribution arises from its power-law
tail, which defines the L\'{e}vy exponent $\gamma=2\nu$ in
Eq.~(\ref{Levyd}). Note that,  again, the generalized average $E_q$
is irrelevant to the  determination of the equilibrium distribution
of the thermometer.

Summing up, our kinetic description of a thermometer in contact 
with a system in a stationary state different from Boltzmann 
equilibrium, suggests that a suitable extension of thermostatistics 
should yield L\'{e}vy distributions as the generalization of the 
Maxwellian velocity distribution, instead of the power-like 
functions of Eq.~(\ref{qexp}). We conclude that the energy 
probabilities derived from Tsallis's thermostatistics do not 
play the role of equilibrium distributions for systems in thermal 
contact. Though this conclusion is not completely general, it 
applies to a wide class of interaction models in the realistic 
situation where a system is put in contact with a thermometer. 
The question remains open as to which form of the entropy and 
which constraints should be used to derive L\'{e}vy distributions 
from a variational principle.

We thank G.~Abramson and I.~Samengo for their critical reading of
the manuscript.


\begin{thebibliography}{00}
\bibitem{not}
Our notation is compatible with the ``renormalized'' Tsallis's
formalism \cite{renorm} and, therefore, differs from Abe and
Rajagopal's.
\end{thebibliography}
\end{document}